\documentclass[%
preprint,%
]{revtex4-1}

\usepackage{graphicx}
\usepackage{dcolumn}
\usepackage{bm}
\usepackage[T1]{fontenc}
\usepackage{textcomp}

\usepackage{mathptmx}
\usepackage{units}
\usepackage{amsmath}
\usepackage{upgreek}
\usepackage{hyperref}
\usepackage{breakurl}

\begin{document}
\setlength{\parindent}{0pt}

\title[Interference of co-propagating Rayleigh and Sezawa Waves observed with micro-focussed Brillouin Light Scattering Spectroscopy]{Interference of co-propagating Rayleigh and Sezawa waves observed with micro-focussed Brillouin Light Scattering Spectroscopy}

\author{Moritz Geilen}
\email{mgeilen@physik.uni-kl.de}
\author{Felix Kohl}
\affiliation{Fachbereich Physik and Landesforschungszentrum OPTIMAS, Technische Universit\"at Kaiserslautern, Germany}
\author{Alexandra Stefanescu}
\author{Alexandru M\"uller}
\affiliation{National Institute for Research and Development in Microtechnologies, Bucharest R-07719, Romania}
\author{Burkard Hillebrands}
\author{Philipp Pirro}
\affiliation{Fachbereich Physik and Landesforschungszentrum OPTIMAS, Technische Universit\"at Kaiserslautern, Germany}

\date{\today}

\begin{abstract}
We use micro-focussed Brillouin light scattering spectroscopy ($\upmu$BLS) to investigate surface acoustic waves (SAWs) in a GaN layer on a Si substrate at GHz frequencies. Furthermore, we discuss the concept of $\upmu$BLS for SAWs and show that the crucial parameters of SAW excitation and propagation can be measured. We investigate a broad range of excitation parameters and observe that Rayleigh and Sezawa waves are excited simultaneously at the same frequency. Spatially resolved measurements of these co-propagating waves show a periodic pattern, which proves their coherent interference. From the periodicity of the spatial phonon patterns, the wavevector difference between the two waves has been identified and compared to the dispersion relation. This concept of co-propagating phonons might be used to produce acoustic or magneto-elastic fields with a time-independent spatial variation similar to the situations realized using counter-propagating waves. However, co-propagating SAW have a well defined direction of the wave vector and thus, posses a finite phonon angular momentum which offers new opportunities, e.g. for angular momentum conversion experiments.
\end{abstract}

\maketitle

\section{\label{sec:Intro} Introduction}
Phonons are the quasiparticles of lattice vibrations and therefore one of the fundamental excitations in a solid body. Of special interest are surface acoustic waves (SAWs), which are bound to the surface of a body. These waves find various applications in industry and research \cite{Delsing2019,Edmonson2004,Mauricio2005}. Since the velocity of acoustic waves is about 100.000 times slower than the speed of light, they are used in miniaturized devices at GHz frequencies. Furthermore, SAWs can be excited in a controlled and efficient manner by so-called interdigital transducers (IDT) on piezoelectric substrates. Therefore, devices based on SAWs are essential building blocks for RF technology and widely used in current wireless devices for telecommunication (LTE and 5G), sensors \cite{Polewczyk2020,Kuszewski2018} or radar systems. Beside these technological applications, SAWs are also subject of fundamental research. Due to the advances in nano-lithography, higher frequencies and smaller wavelengths can be achieved. This opens the possibility to efficiently couple SAWs to other systems (e.g., qubits\cite{Satzinger2018, Manenti2017}, spin waves \cite{Verba2018,Dreher2012} and other hybrid systems \cite{Whiteley2019}). In this context, since SAW can carry phonon angular momentum, the conversion of angular momentum between the phonon and spin system has recently attracted significant attention \cite{Sasaki,Long2018}.\\
Techniques based on IDTs working as sender and receiver offer a very sensitive way to probe the transmitted power. However, this technique provides only limited information about the spacial distribution of the acoustic field in between the IDTs. This field, however, is an important basis for many phenomena. It can be mapped by various methods like atomic force microscopy \cite{Hesjedal2001}, scanning electron microscopy \cite{Eberharter1980} and x-ray measurements \cite{Sauer1999}. In this paper, we want to emphasize micro-focussed Brillouin light scattering spectroscopy ($\upmu$BLS) as an additional method for mapping SAWs in the GHz regime. This method is already well established in the field of magnonics for probing the spacial distribution of spin waves on the micrometer scale \cite{Sebastian2015}.

\section{\label{sec:Methods}Methods}
\begin{figure}[h]
	\begin{center}
		\scalebox{1}{\includegraphics[width=8.0 cm, clip]{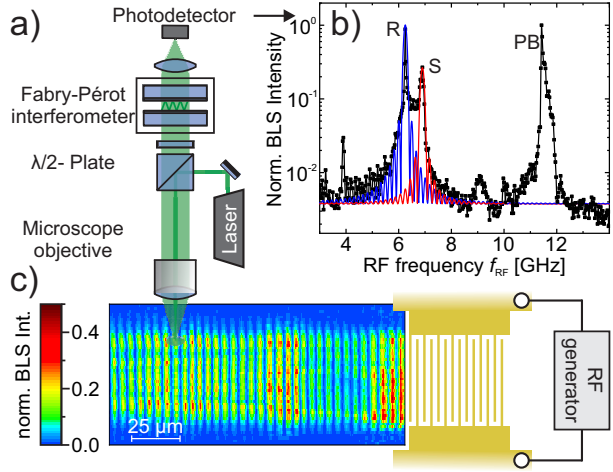}}
	\end{center}
	\caption{\label{fig:fig1} a) schematic BLS setup. The laser light is focussed by a microscope objective on the sample. The polarization of the backscattered light can be rotated by a $\lambda$/2-plate and is analysed by the polarization-sensitive Tandem Fabry-P\'erot interferometer. b) Normalized BLS intensity (logarithmic scale) for different excitation frequencies $f_{\mathrm{RF}}$ applied to the IDT. The square of the calculated excitation efficiency for the Rayleigh mode (blue) and Sezawa mode (red) is shown. c) Two-dimensional time-averaged BLS intensity map of a SAW emitted from an IDT with $f_{\mathrm{RF}}=\unit[6.25]{GHz}$. The SAW form a confined beam with a periodic interference pattern.}
\end{figure}

Brillouin light scattering is the inelastic scattering of photons by quasiparticles such as phonons. During the BLS process, momentum and energy of the scattered photons are changed. Wavevector-resolved BLS spectroscopy is a standard tool to study thermally\cite{Wittkowski2000} and externally excited\cite{Vincent2005,Kruger2004} phonons. By varying the angle of incidence, the probed wavevector can be defined, whereby the dispersion relation can be mapped. However, the spatial resolution is limited by the laser spot size, which is usually some tens of $\upmu$m. Hence, this technique is not suited to investigate phonons on the micro- and nanoscale.\\
For this reason, Carlotti et \textit{al}. \cite{Carlotti2018} emphasized to use micro-focussed BLS, a technique developed in the field of magnonics \cite{Chumak2015}. A scheme of the setup is shown in Fig.\ref{fig:fig1} (a). While the physical principle of the scattering process stays the same, the light of a solid state laser ($\lambda=\unit[532]{nm}$) is strongly focussed onto the sample by a microscope objective. In this study, the objective has a numerical aperture of $NA=0.75$ and a magnification of 100x. This allows a spatial resolution down to $\unit[250]{nm}$ \cite{Sebastian2015}. Due to the focussing, the angle of incidence is no longer well-defined. Therefore, the BLS signal averages over a large wave-vector regime. The shift in frequency of the reflected photons can be measured by a Tandem Fabry-P\'erot interferometer (TFPI). The intensity of the scattered light is proportional to the power of the acoustic wave. Furthermore, the polarization of the backscattered light can be analysed. The polarization entering the interferometer can be chosen by a $\lambda$/2-plate in front of it. Since the interferometer itself is optimized for only one polarization axis, this allows to investigate the polarization of the scattered light. For investigations at the microscale, it is essential to control the position of the laser spot on the sample. Hence, an automatized positioning system combined with microscopic imaging is used. Both the data acquisition by the TFPI and the positioning system were controlled by central programming interfaces developed by THATec Innovation GmbH.\\
The investigated samples consists of single SAW resonators, which have been fabricated on commercial GaN/Si wafer (produced by NTT-AT Japan). Undoped GaN ($\unit[1]{\upmu m}$) is grown on a Si substrate with a $\unit[0.3]{\upmu m}$ buffer layer. This is a so-called "fast on slow" structure, where confined modes, like Sezawa waves, can exist beside the Rayleigh mode. Three different IDT structures have been produced by electron beam lithography and conventional lift-off techniques. Their finger/interdigit widths are $w=\unit[170]{nm},~\unit[200]{nm}~\mathrm{and}~\unit[250]{nm}$. The IDTs are contacted by a PicoProbe connected to an RF generator.

\section{\label{sec:Experiment}Experimental Results}
\begin{figure}[h]
	\begin{center}
		\scalebox{1}{\includegraphics[width=8.0 cm, clip]{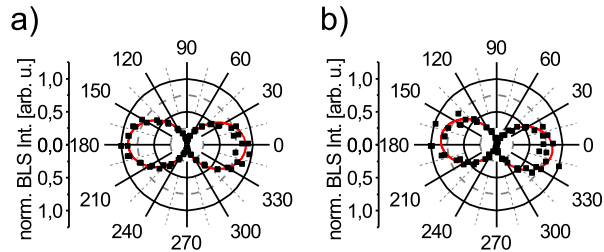}}
	\end{center}
	\caption{\label{fig:fig0} BLS intensity of Rayleigh waves a) and Sezawa waves b) as a function of the angle of the $\lambda$/2-plate in Fig.1. Zero angle corresponds to the reference polarization of elastically scattered light. The polarization of the backscattered light is not changed during the scattering process with Rayleigh waves or Sezawa waves.}
\end{figure}

An RF voltage applied to the IDT generates an alternating strain in the piezoelectric layer, and SAWs can be generated, which propagate away from the IDT. Fig.\ref{fig:fig1} (b) shows the time-averaged BLS intensity close to the IDT with a finger/interdigit width $w=\unit[170]{nm}$ as a function of the RF frequency $f_{\mathrm{RF}}$. The different maxima can be identified as the Rayleigh mode (R) at $\unit[6.25]{GHz}$ and the first Sezawa mode (S) at $\unit[7.0]{GHz}$. Furthermore, a pseudo-bulk mode (PB) has been found at higher frequencies. These three modes can also be observed for the other two devices with different finger/interdigit widths and therefore different wavevectors. The extracted frequencies for all three devices are shown in Fig.\ref{fig:fig3} (a). They correspond well to the dispersion relation that was obtained based on data presented by Mueller et \textit{al}. \cite{Muller2015} for a GaN layer on Si. Moreover, the square of excitation efficiency has been calculated based on the Fourier transform of the IDTs geometry \cite{Royer2000} as a function of the wavevector $k$. Taking into account the dispersion relation, the excitation efficiency can be calculated as a function of the frequency for a given mode. This is shown in blue for the Rayleigh mode and in red for the first Sezawa mode. Both frequencies of the intensity maxima correspond well to the measured data. The noise level of the measurement has been added to the calculated curves. Please note that a absolute comparison of the calculated excitations efficiencies of Rayleigh and Sezawa mode is not possible, since the model does not take into account the strain profile of the IDT, the amplitude profile of the SAW (over the film thickness, respectively) and the surface sensitivity of the BLS. However, for the further discussion, it is important to notice that the excitation efficiency of the Sezawa mode at the Rayleigh mode resonance is not vanishing and vice versa.\\
An additional interesting parameter, which has to be addressed is the polarization of the scattered light. It has been analyzed by a rotation of the $\lambda$/2-plate in combination with the polarization-sensitive TFPI. Rayleigh waves have been excited at $f=\unit[6.25]{GHz}$ and Sezawa waves at $f=\unit[6.9]{GHz}$. As a reference polarization, elastically scattered light from a laser mode has been used. Figure \ref{fig:fig0} shows that the measured BLS intensity follows a $\sin^2$ dependence of the $\lambda$/2-plate angle, indicating that the scattered light is linearly polarized and not changed in its polarization direction. This is in agreement to studies carried out with wavevector-resolved BLS\cite{Carlotti2018}.  

As mentioned above, $\upmu$BLS is a scanning method, which allows to measure the acoustic field at the surface of the sample. As an example, Fig.\ref{fig:fig1} (c) shows the time-averaged BLS intensity of emitted SAWs with a frequency of $\unit[6.25]{GHz}$. Since the aperture of the IDT ($a=\unit[50]{\upmu m}$) is large compared to the wavelength of the SAW ($\lambda=\unit[680]{nm}$), the waves are emitted in a confined beam with no visible diffraction. Hence, the wave can be treated as a plain wave in the following explanation.

\begin{figure}[h]
	\begin{center}
		\scalebox{1}{\includegraphics[width=8.0 cm, clip]{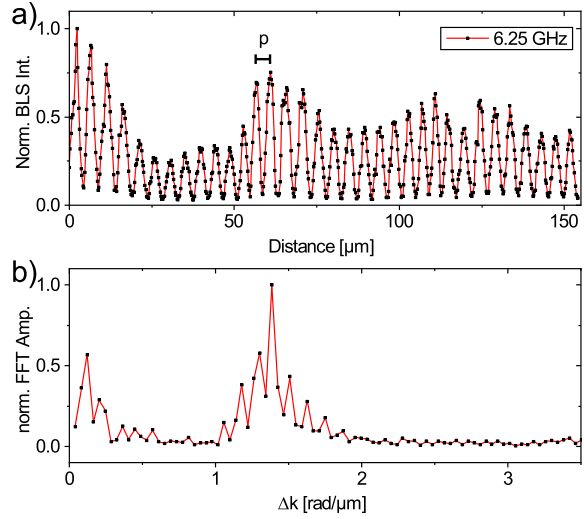}}
	\end{center}
	\caption{\label{fig:fig2} a) Normalized time-averaged BLS intensity for an excitation frequency $f=\unit[6.25]{GHz}$ along the propagation direction extracted from Fig.\ref{fig:fig1}c). b) Normalized FFT of the BLS intensity. The wave-vector difference $\Delta k$ can be extracted from the position of the maximum.}
\end{figure}

It is important to note that the time-averaged BLS intensity oscillates along the propagation direction of the SAW with a periodicity $p=\unit[4.59]{\upmu m}$. For better visibility of this phenomenon Fig.\ref{fig:fig2} (a) shows the BLS intensity summed over the width of the measured area. This periodic oscillation in intensity can be explained by the interference of co-propagating Rayleigh and Sezawa waves with the same frequency, but different wavevectors. This is possible due to the finite length of the IDT and the associated linewidth in k-space (see Fig~\ref{fig:fig1}~(b)).\\
We assume two plane waves $\Psi_i$:
\begin{equation}
\Psi_i = A_i \cos(k_{i}x-\omega t).
\label{eqn:eqn1}
\end{equation}
The BLS intensity is proportional to the SAW intensity averaged over one oscillation period, which is expressed by:
\begin{equation}
\label{eqn:eqn3}
\begin{split}
I(x) & =\vert{\int{\Psi_R(x,t)+\Psi_S(x,t)\ dt }}\vert^2\\
& =A_R^2+A_S^2+{2{A_R A_S}}\cos[\Delta k x]
\end{split}
\end{equation}
with
\begin{equation}
\Delta k = k_R-k_S = {{2\pi}\over p}.
\label{eqn:eqn4}
\end{equation}
From the amplitude of the observed modulation the ratio between the amplitude of Rayleigh and Sezawa wave can be estimated to $A_S/A_R \approx 0.5$. This value is larger than one might estimate from the simple model shown in Fig. 1. However, as already explained before, the different excitation efficiencies together with the different BLS scattering cross sections for the two modes prevents a direct comparison. \\

\begin{figure}[h]
	\begin{center}
		\scalebox{1}{\includegraphics[width=8.0 cm, clip]{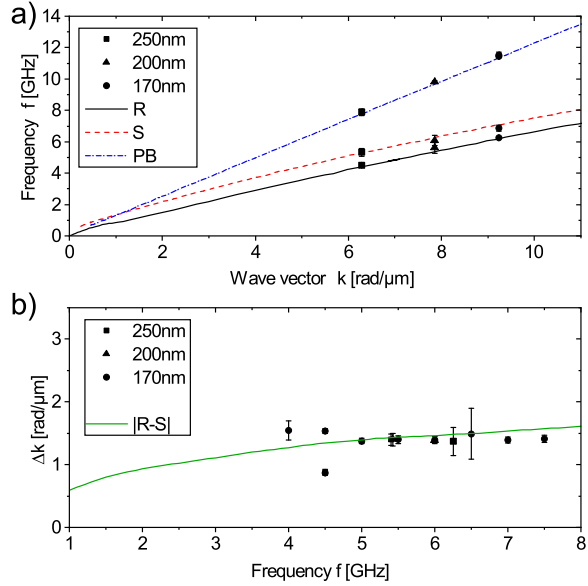}}
	\end{center}
	\caption{\label{fig:fig3} a) Dispersion relation for SAWs in a $\unit[1.3]{\upmu m}$ thick GaN layer on Si substrate. From the BLS measurements for different excitation frequencies the maxima are evaluated and the positions are shown (black dots) for all tested devices (see Fig. \ref{fig:fig2} (b)). b) The wave-vector difference $\Delta k$ for a given frequency between Rayleigh and Sezawa mode has been determined by BLS measurements along the propagation direction. The predicted wave-vector differences based on the dispersion relation are represented in green color.}
\end{figure}

The wave-vector difference $\Delta k$ has been extracted from the Fourier transform of the BLS intensity along the propagation direction, as shown in Fig.\ref{fig:fig2} (b). To confirm the interpretation of the two co-propagating and interfering waves, additional measurements at different excitation frequencies $f_{\mathrm{RF}}$ have been performed. Therefore, the BLS intensity was mapped on a line along the propagation direction and the beating pattern was analyzed like shown above. All measurements showed similar oscillations. The resulting wave-vector differences are shown in Fig.\ref{fig:fig3} (b). In the investigated frequency range the wave-vector difference $\Delta k$ is nearly constant, which is in a good agreement with the prediction based on the dispersion relation of the GaN-Si system. While for small wavevectors the slope for the Rayleigh and the Sezawa mode varies it is comparable to waves with frequencies above $\unit[2]{GHz}$. Deviations can be attributed to the fact that the value of the wave-vector difference is very sensitive to small changes of dispersion.\\

\section{\label{sec:Conclusion}Conclusion}
In this article we have used micro-focussed Brillouin light scattering spectroscopy to investigate the excitation and propagation of SAWs in a GaN layer on a Si substrate. In the first part, the excitation efficiency of the IDT and dispersion relation has been verified. We have found that Rayleigh and Sezawa waves are excited at the same frequency by an IDT structure. These waves interfere coherently while they propagate and their time-averaged intensity is spatially varying with a periodicity determined by their wavevector difference at the excitation frequency. This pattern offers the opportunity to have a time-independent, space-modulated acoustic field with well defined wave vector direction.

\begin{acknowledgments}
Financial support by the EU Horizon 2020 research and innovation program within the CHIRON project (contract no. 801055) is gratefully acknowledged. 
\end{acknowledgments}

%
%
%
\bibliographystyle{apsrev4-1}
\bibliography{InterferenceofRayleighandSezawawavesversion_arXiv}



\end{document}